\documentclass[preprint]{aastex}
\def\magsqa{mag/$\prime\prime^2$}

\def\mucfit{$\mu_0^{\rm fit}$}
\def\kms{km s$^{-1}$}
\usepackage{emulateapj5}
\newcounter{nfig}
\begin{document}
\singlespace
\pagenumbering{arabic}               

\title{The $z=0$ Galaxy Luminosity Function: I. Techniques for Identification of
Dwarf Galaxies at $\sim 10$ Mpc.\altaffilmark{1}}

\author{Kathleen Flint, Anne J. Metevier, Michael Bolte\altaffilmark{2}}
\affil{UCO/Lick Observatory, University of California, Santa Cruz, CA 95064}
\affil{flint@ucolick.org, anne@ucolick.org, bolte@ucolick.org}

\and

\author{Claudia Mendes de Oliveira\altaffilmark{2}}
\affil{Instituto Astr\^{o}nomico e Geof\'{i}sico (IAG), Av. Miguel 
Stefano 4200, CEP: 04301-904,\\ S\~{a}o Paulo, Brazil}
\affil{oliveira@iagusp.usp.br}

\altaffiltext{1}{Lick Bulletin No. 1399}
\altaffiltext{2}{Visiting Astronomer, Kitt Peak National Observatory,
National Optical Astronomy Observatories, which is operated by the
Association of Universities for Research in Astronomy (AURA), Inc.,
under cooperative agreement with the National Science Foundation.}

\shorttitle{The $z=0$ Galaxy Luminosity Function I.}
\shortauthors{Flint et al.}
\slugcomment{Accepted for publication in the Supplement Series of the Astrophysical Journal 2001}

\begin{abstract}

We present a program to study the galaxy luminosity
function (GLF) of the Leo I and Coma I Groups at $\sim10$ Mpc. 
We have surveyed over seven square degrees in  Leo I
and $\sim$11 square degrees in  Coma I.  In this paper, we
detail the method we have developed and
implemented for identifying on morphological grounds
low-surface-brightness, $M_R < -10$ dwarf 
galaxies at a distance of 10 Mpc. We also describe extensive
Montecarlo simulations of artificial galaxies which we use to tune our
detection algorithms and evaluate our detection efficiency and
parameter recovery as a function of $\mu_R(0)$ and $R_T$.  We find for
a sub-set of our Leo I data that at the 90\% completeness level we
can detect dwarfs comparable to Antlia and Sculptor.  
Finally, we describe preliminary follow-up observations which confirm
we are detecting dwarf 
spheroidals in Leo I at 10 Mpc.

\end{abstract}

\keywords{galaxies:dwarf---galaxies:luminosity function, mass
function---techniques: image processing}

\section{Introduction}

Knowledge of the galaxy luminosity function (GLF) is of fundamental
importance to extragalactic astronomy and cosmology.  Studies of the
GLF in different density environments and at different redshifts
provide us with the clues to answer questions about how galaxies 
were formed and how their properties and number counts have evolved.
The {\it local} GLF is particularly important for two reasons.  First, it is 
virtually impossible to observe faint, low-surface-brightness objects at
cosmologically interesting distances.  Our understanding of the
luminosity density of the Universe must therefore come from
observations of distant, high-luminosity galaxies in combination with
an extrapolation to fainter objects based on measurements of the local
GLF (while keeping in mind possible evolutionary effects). Second, in
order to interpret faint galaxy counts in the context of evolution of galaxy
properties and to map distant populations to local counterparts, the
$z=0$ GLF is necessary for establishing a baseline of 
current-day properties and space densities for galaxies of all luminosities.

\subsection{A Brief History of GLF Measurements}

Because of the importance of the GLF, deriving it in different
environments has been the subject of many investigations in the past
two decades. There have been many approaches taken to estimate the GLF
for the field, in the vicinity of luminous galaxies, in groups, in the
nearby Virgo and Fornax clusters and in more-distant rich clusters.
We make no attempt to give a complete overview of the field here.
\citet{pri99} have a concise summary of recent
studies of the GLF in different environments and \citet{tre98b}
reviews GLF determinations in somewhat more detail. 
\citet{imp97}
also review recent determinations of the GLF and have extensive
discussions of issues specifically related to selection effects in
studies of low-surface-brightness galaxies. 

We use the standard Schechter function \citep{sch76} to describe the
GLF.  The number of
galaxies  per unit volume ($\phi$) at luminosity $L \pm dL/2$ is given by 
\begin{equation}
\phi (L)dL = \phi^{*} (L/L^{*})^{\alpha} e^{-L/L^{*}} dL
\end{equation}
where $L^*$ is the luminosity of galaxies at the `break'
between the exponential form at high $L$ and power law  at low
$L$ and  $\phi^{*}$ is the space-density normalization.
The characterization of the GLF to the lowest luminosities 
--- dwarf spheroidal (dSph) galaxies like Draco at $M_R\sim -9$ ---
is likely near complete only for the complement of companions to the
Milky Way. Yet for the Local Group (LG), M31 dSph galaxies are still
being discovered 
\citep[{\em e.g.} ][]{arm99} as are intra-group dSph 
\citep[{\em e.g.} ][]{lav92,kar98}.
With the currently known members, 
the Schechter
parameters for the LG are $M^*\sim
-21; \alpha \sim -1.1$ for the $R$-band \citep{pri99}.

Beyond the LG, GLF studies have been carried out for the field
based on redshift catalogues and careful analysis of the selection
function for the catalogues \citep[{\em e.g.}
][]{lov92,mar94,lin96,zuc97,mar98}, in nearby clusters and groups  
\citep[{\em e.g.} ][]{san85,fs90,zab00} and in more
distant clusters \citep[{\em e.g.} ][]{ber95,tre98a,deP98}.  
Despite this rich history of measurements, there is no consensus on
the nature of the faint-end slope of the GLF and how it may depend
upon environment and morphology.  Recent studies in 
clusters have measured a steep faint-end slope 
\citep[{\em c.f.} ][]{tre98a,hra00}, suggesting that higher density
regions have a higher dwarf-to-giant ratio, yet some studies of the
low-density field GLF have found similarly steep results \citep[{\em
c.f.} ][]{zuc97,lov97}.  There is also evidence that the faint end
varies with Hubble-type \citep[{\em e.g.} ][]{jer97,mar98}, color
\citep[{\em e.g.} ][]{met98}, and emission-line strength \citep[{\em
e.g.} ][]{chr00}, and in some cases is contrary to what is found in
the LG for similar environments. 
For
example, the study of \citet{mar98} includes both LG-density groups and
the field and reports a relatively flat
GLF for elliptical and spiral galaxies and a steeper GLF
for peculiar and irregular galaxies, whereas in 
the LG the irregular GLF is relatively flat.

The majority of the studies outside the LG to date are
restricted to galaxies brighter than $M_R\sim -16$, which only begins
to probe the faint end of the GLF. Exceptions to this
are (curiously) some of the studies of relatively distant clusters
where the claimed low-luminosity limits are $M_R\sim -12$ \citep[{\em
e.g.} Coma,][]{ber95}. These studies have generally taken a
statistical approach in subtracting off background and foreground
contamination and have pushed the faint limits of their GLFs to the
limits of the photometry.  Potential problems with this approach
(particularly at faint levels where the surface density of distant
galaxies is high) are projection effects \citep{val00} and  the existence of large-scale structure giving rise
to significant variations in the background counts as a function of
position.  Most studies concentrating on nearby groups and clusters
(necessarily extending over large areas on the sky) have identified
low-surface-brightness (LSB) dwarfs on photographic plates,
usually followed up with CCD observations ({\em e.g.} in Virgo
and M81; Impey, Bothun, \& Malin 1988, Caldwell et al. 1998).
Only
recently have wide-field CCDs been available to cover similarly
large expanses of the sky \citep[{\em c.f.} ][]{car01}.

\subsection{A New Survey}

We have undertaken a program to measure the GLF to very faint levels in
nearby ($\sim10$ Mpc $<$ d $<$ 15 Mpc) galaxy groups using $R$-band
MOSAIC CCD images
for the initial survey and a number of different techniques to
establish group membership of candidates on a galaxy-by-galaxy basis.
Our primary goal is to measure the $z=0$ GLF to luminosity and
completeness limits similar to that achieved in the LG. This
will allow a test of the variation of the GLF between groups, decrease
the shot noise in the GLF by combining samples or simply having a larger
sample of galaxies than is present in the LG.  Furthermore, it will provide
a large sample of low-luminosity galaxies that are near enough to be amenable to
fairly detailed studies of individual galaxy properties.

At these distances, dwarf galaxies, at least those represented in the
LG, have apparent sizes $>10^{\prime\prime}$ at our
surface-brightness limit (50\% completeness) of $\mu_R\sim 26$
mag/$\prime\prime^2$. It may therefore be possible to identify
dwarf-galaxy members purely on morphological grounds. 
The imaging data of our survey were taken with 
the MOSAIC camera on the KPNO 0.9-meter telescope (described more
fully in \S 2) with the engineering-grade CCDs.  Due to 
the poor blue response of these original chips, we imaged our survey
with the Cousins $R$ filter.
Figure \arabic{nfig} shows
the positions of LG dwarfs as they would be seen at 10 Mpc in
a plot of total magnitude vs. central surface brightness.
LG dwarf properties are taken from \citet{mat98} and references
therein, adopting $\langle B - R \rangle \simeq 1.3$.
The dashed
lines show apparent sizes of galaxies with exponential profiles in
this plane at a limiting 
surface brightness isophote of $\mu_R=26.7$ mag/${\prime\prime}^2$. Most
of the LG galaxies have diameters at this isophote level that
are greater than 10$^{\prime\prime}$.

We describe in detail our procedures for identifying intrinsically faint
dwarf galaxies in the following sections. We note here that
there are not very many types of objects known
with the low central surface brightness level of a typical
dSph. Distant 
high-luminosity galaxies are subject to $(1+z)^4$ surface brightness
dimming and enter into our regime of interest, but these objects will
typically be apparently small --- a few arcseconds or less at our
isophotal limit. Background large low-surface-brightness galaxies are
potential sources of confusion with our method \citep{dal97} as are very distant galaxy clusters for 
which the main signature is a large, low-surface brightness enhancement
due to the overlapping envelopes of the cluster galaxies
\citep{zar97}.  However, with a somewhat conservative isophotal
diameter limit of 10\arcsec\ these objects should not be a significant
contaminant. 
We will carry out group-membership studies (see \S5.2) for
at least a 
subset of our morphologically-identified candidate members.

The initial groups in this survey are the Leo I group \citep[NBGC
15-01 or LGG 
217, containing NGC 3379 and NGC 3368;][]{tul88,gar93} and the Coma I `cloud'
\citep[NBGC 14-01 or LGG 279, containing NGC 4278 and NGC
4314;][]{tul88,gar93}. In this paper we describe our
algorithms for searching the frames for candidate group members,
techniques for deriving photometric and structural properties of the
candidates and simulations used to tune the search technique parameters
and evaluate completeness and measuring errors. We also present
examples of the initial findings. In a subsequent paper we will present
the details of the observations, our candidate lists,
charts, and the GLF.

\section{Survey Imaging}

This program begins with direct imaging using the MOSAIC Camera on the
0.9-meter telescope at KPNO.  This camera gives a field of view
$1^\circ$ on a side and is comprised of eight $2048 \times 4096$ CCDs,
with sampling  (unbinned) of 0.42 arcsec/pixel. The details of the
observations for each group will be presented in the subsequent data papers.
 In short, nine  fields were observed in the Leo
Group and twelve  fields in the Coma I Group. For most of
the Leo Group observations the original engineering-grade CCDs 
of the MOSAIC were used. For each field, five 900-second exposures were
made through the Cousins $R$ filter using the standard dithering
routine with telescope offsets of $\sim 
40^{\prime\prime}$ between the images. There were significant gaps
between the CCDs ($\sim 21\arcsec$ N-S, $\sim 15\arcsec$ E-W ) in the MOSAIC and numerous cosmetic flaws in the
engineering-grade CCDs. The purpose of the in-field dithering was to
allow removal of these artifacts in the process of combining the five
frames into a single image. In retrospect, we should have had a larger
number of dithered pointings to better remove artifacts and improve our
ability to flat-field on intermediate scales.

\addtocounter{nfig}{+1}
Our procedures for bias and dark subtraction, flat-fielding and
combining the frames followed the steps detailed in the {\it Guide to the
MOSAIC Data Handling System} \citep{msc98}.  To carry out this initial
image processing we used 
the {\tt MSCRED}  package within IRAF\footnote{IRAF is distributed by the National Optical Astronomy Observatories,
which are operated by the Association of Universities for Research
in Astronomy, Inc., under cooperative agreement with the National
Science Foundation.}.  The preliminary processing of
MOSAIC frames is somewhat complicated and will be discussed in detail
in a subsequent paper. To do photometric calibration of the stacked MOSAIC
images, we independently observed our Leo I and Coma I
fields in $B$ and $R$ with the Lick Observatory 1-m Nickel telescope.
The calibration is good to $\pm0.04$ mag in $R$, but when we include 
the uncertainty in the color-term from our estimates of the $B-R$
colors of the detected objects, the final galaxy photometric error is
$\pm0.07$ mag. 
Figure \arabic{nfig} highlights one of the nine
fields imaged in the Leo I Group, with the 0.9-m fields overlaid upon
a Digital Sky Survey image\footnote{Based on photographic data
obtained using Oschin Schmidt Telescope 
on Palomar Mountain.  The Palomar Observatory Sky Survey was funded
by the National Geographic Society.  The Oschin Schmidt Telescope is
operated by the California Institute of Technology and Palomar
Observatory.  The plates were processed into the present compressed
digital format with their permission.  The Digitized Sky Survey was
produced at the Space Telescope Science Institute (ST ScI) under
U. S. Government grant NAG W-2166.}. 
The $\sim3\arcdeg \times \sim3\arcdeg$ imaging area encompasses the
1\fdg1 virial radius of the group \citep{tul87}. We analyze the
indicated field (Field 7) to demonstrate
our search and simulation algorithms. Our coverage of the Coma I
group, while larger in extent ($\sim11$ deg$^2$) covers a smaller
fraction of the extensive group ($>25$ deg$^2$).

\section{The Analysis Pipeline: Detecting Faint Galaxies}

We have developed a three-step process for detecting dwarf galaxies at
10 Mpc.  With the first, we detect high-surface-brightness (HSB) galaxies
through a standard $k$-$\sigma$\ thresholding technique using the 
SExtractor object
detection and classification software \citep{ber96}.  In the second
step, we use a search method we have optimized to find LG-like dwarf galaxies
that would typically fall below the detection limits of most standard
methods. 
We use a matched-filter technique for which the HSB galaxies and bright stars are masked, and the
resulting image is convolved with an exponential 
filter representative of the structure of a LG dwarf at a distance of
10 Mpc.
This second step of our detection process is based in part on the
technique developed by \citet{dal95} to detect large
low-surface-brightness (LSB) galaxies in the field from drift-scan
data. In the third step of our analysis pipeline we characterize 
the detections and create a reliable candidate list from which to do
further analysis and observations.
In this section, we describe our detection methods  
and illustrate them in Figures \addtocounter{nfig}{1}\arabic{nfig}
-- \addtocounter{nfig}{3}\arabic{nfig}.

\addtocounter{nfig}{-3}

\subsection{Step 1: Traditional Detection Method}

Before object detection, each image was pre-processed to remove the
brightest objects from the image.   We modeled the largest,
most luminous galaxies in the field using the IRAF task {\tt ELLIPSE}, and
subtracted off their elliptical light components.  This revealed 
objects near the line of sight of the large galaxies, as well as
remaining structure in the galaxies.  We then removed the outer light from all
stars brighter than $m_{R} \simeq
10$ --- a somewhat arbitrary cut that removed the largest stars with
diffraction spikes, but didn't present an unreasonable number of stars
to fit by hand.  The saturated centers and diffraction spikes of the stars were
masked with an IRAF routine that flags objects brighter than a
specified  threshold, and these masked objects were then replaced with a 
median-smoothed approximation of the background from SExtractor plus
artificial noise added with the IRAF task {\tt MKNOISE}. 
We modeled the resulting smoothed star image with {\tt ELLIPSE}, and
subtracted the modeled light from the original, unmasked image.

Once we rid our image of diffuse light from bright objects, we ran 
SExtractor in `single image mode', where detection, photometry, and shape
measurements were all done on the same input image.  We set a 70-pixel
minimum contiguous area for detection and a detection threshold of
1.5-$\sigma_{sky}$ above the sky.  We required SExtractor to create noise
maps of the image internally, and weight detections according to the
noise characteristics of the section of the image where the detection
was made.  For star/galaxy separation, we used the object classifier of
SExtractor, {\tt CLASS\_STAR}, and kept only objects with 
{\tt CLASS\_STAR} $\leq 0.1$ and semi-major axis of the SExtractor
ellipse fit, {\tt A\_WORLD} $> 1$\arcsec, which corresponds roughly to
an isophotal radius of 3\arcsec\ (we found that for 
the  smallest 
objects, SExtractor could no longer distinguish stars from galaxies).
With the remaining objects, we fit radial profiles with {\tt ELLIPSE}
in IRAF,
inputting the measured SExtractor shape parameters as initial fit
values.  
We fit the one-dimensional isophotal profile of each galaxy
output from {\tt ELLIPSE}, assuming an exponential profile.  
We then characterize our detections from the traditional method by
adopting the central surface 
brightness as extrapolated from the radial profile fit, and 
the magnitude as measured by SExtractor, {\tt MAG\_BEST}.
Of these objects we retain those with a limiting isophotal size at
$\mu_{lim} = 26.7$ of $\Theta_{lim} > 10\arcsec$ assuming an
exponential profile with zero ellipticity \citep{all79}. 
Further object selection was done
through morphological membership selection on a galaxy-by-galaxy
basis, examining each object by eye  (described in \S 5). 
We used simulations 
(see \S\addtocounter{section}{+1}\arabic{section}) to tune all
detection parameters used in this step.  
\addtocounter{section}{-1}

To demonstrate the first step of our galaxy detection process,
we show in Figure \arabic{nfig} a $15\arcmin \times 15\arcmin$  sub-section of
our Field 7 
image, on which we have placed simulated images of LG
galaxies as they would appear at a distance of 10
Mpc (the galaxy simulation procedure is described in 
\addtocounter{section}{+1}\S\arabic{section}).\addtocounter{section}{-1} 
We have modeled all dSphs as exponentials. 
Note that at this distance Leo II, Tucana, And V, and Carina are extremely
difficult to detect by 
eye.  In Figure \addtocounter{nfig}{+1}\arabic{nfig}, we show the
results of Step 1, where an aperture is drawn around every object
detected that is also classified as a galaxy by SExtractor ({\tt
CLASS\_STAR} $\leq 0.1$, {\tt A\_WORLD} $>$1\arcsec).  Many of
the artificial LG dwarfs are detected; however, the
lowest-surface-brightness objects, And II, Tucana, And V and Carina,
are fainter than the 1.5-$\sigma_{sky}$ detection threshold.

\subsection{Step 2: Optimized Detection Method}
With the second method, we essentially removed all objects detected by
the first 
method, and convolved the remaining background with a smoothing kernel
to bring out LSB features.  To remove all bright objects, we 
 masked all HSB objects above a
2-$\sigma_{sky}$ surface-brightness threshold.  We replaced the
masked pixels in our 
field with a single background value plus noise added with {\tt
MKNOISE} (appropriate to the field).
We then convolved the masked image, using the IRAF task {\tt
CONVOLVE}, with a kernel optimized to enhance 
fluctuations of the size and shape of LG-sized dwarf galaxies at 10
Mpc.  In 
this example we used an  exponential kernel with a 5$\arcsec$ scale
length,  as dSphs are known to be reasonably
well-fit by exponential profiles \citep[$I=I_0 e^{-(r/\alpha)}$; 
{\em c.f.}~][]{ fab83,vad94}, and 5\arcsec\ is a characteristic
exponential scale length for LG dSphs if moved to 10 Mpc \citep{mat98}. 
We find, however, that our detection rate is not overly sensitive to
the shape of the convolving kernel.

We ran SExtractor in  `double image mode' on the resulting convolved
images, where the detection and shape measurements
were performed on the convolved image, but the corresponding
photometric measurements were done on the original image.  We set 
the detection limit to require a minimum area of
100 pixels above a 5.5-$\sigma_{sky}$ surface brightness threshold.
The simulations in 
\addtocounter{section}{+1}\S\arabic{section}\addtocounter{section}{-1} 
were used to optimize these parameters.  
SExtractor's photometry in
double-image mode, however, tends to underestimate magnitudes and
sometimes central surface brightnesses due to its inability to
de-blend neighbor objects on the second (measurement) image.
De-blending occurs here on the convolved image, where bright neighbors
have been masked and so not included in de-blending.
For final parameter measurement we
fit radial profiles 
for each optimized detection, as
discussed below. 

There are a small number of spurious detections from this step, as the
convolution process not only enhances features of a certain scale but also
smoothes them in such a way that SExtractor's neural network training
for {\tt CLASS\_STAR} can be fooled.  However, most brightness
enhancements detected in the convolved image can be culled by hand, as
most of these false detections are obviously 
poorly masked stars and brighter galaxies.  An
example of such detections  
can be seen in \addtocounter{nfig}{+1}Figure \arabic{nfig}, where  we
show the masked, convolved image that corresponds to
Figure \addtocounter{nfig}{-2}\arabic{nfig}\addtocounter{nfig}{2}.
Here, all SExtractor detections from Step 2 are indicated: the obvious
masking remnants  detected are marked with crosses, and the
candidate detections are marked with elliptical apertures.

In this example, all artificial LG  dwarfs are detected, except Leo II.
Note the large size of the elliptical apertures, which are  
roughly equivalent to the isophotal radii, for the input galaxy 
detections.  We use the large apparent size of such dwarf
galaxies to  
discriminate against more distant and typically apparently smaller galaxies. 

One challenge for this 
search method is that the masking process may
mask the higher-surface-brightness centers of the sort of very faint
and extended galaxies we hope to detect.  The convolution process may
then smooth the 
remaining light, but the resulting convolved object will have a
fainter peak luminosity and perhaps fall below the detection
threshold.   An example of this in Figure \arabic{nfig} is Leo II,
which was detected in Figure
\addtocounter{nfig}{-1}\arabic{nfig}\addtocounter{nfig}{+1}, 
but is masked and undetected by the optimized method.  One possible
problem would be the case where an object falls just 
below the threshold of the traditional method, but is also masked and
overlooked by the optimized method.  To guard against these losses,
we set the masking threshold (2-$\sigma_{sky}$) to be somewhat higher
than the detection threshold for the 
traditional method (1.5-$\sigma_{sky}$), so that even though some objects
will be partially masked in the optimized method, they are likely to have
already been detected in Step 1.  In this way, we have a
number of intermediate-surface-brightness objects that are detected
twice, but we minimize the chances of complete non-detections of such objects in this
middle ground.

Objects detected via both methods, or by the optimized method alone,
have an extended LSB component that falls below the traditional
surface brightness threshold.  For these objects, the traditional
detection aperture misses this extended light, underestimating 
the total flux.  Furthermore, the SExtractor magnitude measured by the
optimized method tends to be skewed by bright neighbors that may fall
within the measurement aperture, as mentioned earlier. 
We avoid these problems by
measuring the isophotal radial profile with {\tt ELLIPSE}, again
fitting the profile with an exponential.  We then characterize these
detections with the fitted central surface
brightness, $\mu_0^{\rm fit}$, as in the traditional method, but 
for the magnitude, we adopt the total magnitude calculated analytically
from \mucfit\ and the 
fitted exponential scale length, $\alpha^{\rm fit}$.

\addtocounter{nfig}{+1}
In Figure \arabic{nfig}, we plot the 
example detections from Figures
\addtocounter{nfig}{-2}\arabic{nfig} and
\addtocounter{nfig}{+1}\arabic{nfig}\addtocounter{nfig}{+1}
on the $R_T\ {\rm vs.}\ \mu_0$ plane.
Here and in the following figures, $R_T$ is the total magnitude as
measured by each detection method.
  Detections from the traditional method
are indicated as crosses, where we plot the total magnitude measured
by SExtractor ({\tt MAG\_BEST}) and the central 
surface brightness parameters from the fitted profile.  As in seen in
\S4.3, for these higher-surface-brightness galaxies, the profile fit
does a good job estimating the central surface brightness, but
uncertainties in the slope of the profile yield larger uncertainties
in the calculated total magnitude than we see in the SExtractor
magnitude measurement.  We plot optimized detections as stars, where
$R_T$ is now the total magnitude calculated analytically from
\mucfit\ and the fitted exponential scale length, $\alpha^{\rm fit}$. 
The 
artificial input galaxies that were successfully detected are circled.  
Even though all the included objects are detected, Carina and And V
represent the very limits of our detection method as will be
seen in the following section. As
mentioned previously, the proximity of the Leo I and Coma I groups
makes dwarf galaxies large in angular size.  This can be seen in Figure
\arabic{nfig} in the 
separation of smaller, more compact 
background objects (the dense concentration of crosses centered
around $R_T \sim 19.5, \mu_0 \sim 22$) and LG galaxy properties.
Membership characterization will be discussed more fully in \S 5.  The
input galaxies (circled objects) all are measured with 
parameters comparable to their input values and so fall along the main
locus of LG-like galaxies. Our ability to measure
these parameters is explored with the simulations, as described in the
next section.

\section{Simulations}
Two factors affect our ability to measure the GLF: (1) contamination
from non-group members and artifacts, 
such as noise peaks or flat-fielding errors
mistakenly selected through the optimized detection method, and (2)
incompleteness in our detection method.  We
address the first on a galaxy-by-galaxy basis, guided by measured
properties, morphology, and follow-up observations.  The second we
address  via Montecarlo simulations in which
we added artificial galaxies to the data frames and attempted to recover
them with our detection processes described in
\addtocounter{section}{-1}\S\arabic{section}.  
\addtocounter{section}{1}  The final
luminosity function will be membership-corrected for each of these
effects. The final galaxy counts are
the observed counts of member galaxies, $n_{obs} = n_{obs}(R_T,\mu_0)$,
weighted 
by the detection efficiency as determined from the
simulations, $f(R_T,\mu_0)$, where $R_T$ and $\mu_0$ are the input
total magnitude and central surface brightness, respectively, of the
artificial galaxies.
Thus, the final galaxy counts for the GLF will be 
$N(R_T,\mu_0)$ = $n_{obs}/f$. 
We present a subset of these simulations here; the full simulations
results will be presented in a subsequent paper.

\subsection{Simulation Procedure}

In order to understand the complex selection function and to tune our
detection parameters for both  steps in our detection method we
have carried out extensive artificial-galaxy simulations.
We create artificial galaxies covering the full range of the LG-dwarf
magnitudes and central surface brightnesses (as seen at 10 Mpc), degrade
them by the seeing and noise 
properties  of the real data, add them
to the real data frames and then detect them by the 
 method described in
\addtocounter{section}{-1}\S\arabic{section}\addtocounter{section}{+1}.
We calculate the detection efficiency as $ f(R_T,\mu_0) =
n_{rec}/n_{add}$, where $n_{rec}$ is the number of galaxies recovered
and $n_{add}$ is the number of galaxies added for a given bin.
 The artificial galaxies populate the $R_T\ vs.\ \mu_0$
plane spanning the ranges $14 \leq R_T \leq
22$, $22 \leq \mu_0 \leq 26$, roughly corresponding to the range 
wherein we would see LG-counterpart dSphs.

The artificial galaxies we created with 
the IRAF task {\tt MKOBJECT} in the same manner as those shown in
Figure 3.  With {\tt MKOBJECT} we input the
magnitude zeropoint (which includes the color term for an adopted
dwarf color of $B - R = 1.3$), read-noise, and seeing of our
observations, 
and we specify an  exponential radial profile, total
magnitude, characteristic scale length, axial ratio, and position
angle.  The routine approximates the profile analytically and
artificially truncates the profile at a cut-off radius where the
profile intensity is $10^{-5}$ times the peak intensity ($\sim12.5$
mag below the peak).  The dwarf galaxies we simulate are not significantly affected
by this truncated dynamic range. Again, we adopt 
a purely exponential profile as it is known to be a good model for 
faint dSph galaxies  and we include  
 a uniform range of axial ratios to model the galaxies'
intrinsic ellipticities. LG compact dwarf ellipticals like M32 follow a
de Vaucouleurs $R^{1/4}$-law, but their high central surface
brightnesses (as can be seen by M32's position in Figure 1) will make
them easy to detect at these distances. Thus, we do not include these
profile types in our simulations. 
There have long been discussions in
the literature comparing the three-parameter King profile with the
two-parameter exponential profile as the most appropriate for
early-type dwarfs (see Ferguson \& Binggeli 1994 and references
therein), with exponentials being very similar to a King profile with
large core radius and low concentration. Fainter dwarfs ($M_R \lesssim -17$)
appear to be well-fit by an exponential at all radii yet  brighter
($M_R \lesssim -17$) dwarfs, for which we see some profile flattening
({\em e.g.} 
Fornax dwarfs; Bothun, Impey, \& Malin 1991)
or excesses in the inner regions \citep{cal87},
benefit from the more flexible King profile.
 We adopt an exponential profile for our fake galaxies for several
reasons.  First, a two-component model simplifies the parameter space
we explore with our simulations.  Second, 
deviations between these profiles in the core regions are minimized by the
seeing convolution of our simulations at 10 Mpc.  Finally, our
simulations are most critical for the lowest luminosity regime ($M_R >
-14$), where the exponential profile is found to be the most prevalent.

For these simulations we neglected any internal structure, patchiness
due to dust or HII regions and thus internal absorption.  For dSph
galaxies, this is appropriate.  To a first approximation, this also
tells us something about our $R$-band detection efficiency for dwarf
irregulars, which typically have smoother and more extended disk of
old, red stars underlying a younger and bluer stellar population that
is well-fit by an exponential \citep{irw95}.  However, in order to
fully probe the effects of substructure on detectability, we will add
high signal-to-noise images of real LG dwarf irregulars ({\em e.g.} GR
8, IC 10) to our frames, scaled to a distance of 10 Mpc with comparable
noise and seeing to our data, and will attempt to detect them with our
method.  This will be included in a subsequent paper.

We have randomly generated galaxy parameters uniformly in bins of size
0.5 mag/$\prime\prime^2$ in $\mu_0$\ and 0.5 magnitudes in $R_T$,
according to the appropriate radial profile.  We then randomly
distributed each galaxy in the image frame with a spatial density that
allowed for a possible $\lesssim1\% $ overlap of the area in input
objects (assuming a maximum isophotal size  at the 27
mag/$\prime\prime^2$\  isophote).  Preliminary simulations were first
performed by adding exponential-profile galaxies with $17 \leq R_T
\leq  22$, $1\farcs6 \leq \alpha \leq 6\farcs3$ to a purely artificial
noise frame, mimicking the noise characteristics of the CCD, in order
to tune the simulation and detection algorithms.  For the traditional
method, we tuned the detection parameters solely to maximize the
probability of detection.  We found that a detection threshold of
1.5-$\sigma_{sky}$ vs. 2-$\sigma_{sky}$ made little difference, and so
we chose the more generous threshold.  We reduced the minimum detection
area ({\tt DETECT\_MINAREA}) based on the size of artificial LG
galaxies moved to 10 Mpc, adopting the minimum isophotal area above the
detection threshold of the faintest LG-like galaxies we expect to
detect with the traditional method.  Lastly, these simulations also
helped us to tune the background mesh sizes ({\tt BACK\_SIZE}), which
governs how large a background area is effectively smoothed.  We found
the size did not significantly change  the detection efficiencies.

The optimized detection parameters were also tuned from simulations
with these purely artificial noise frames, which we found to be
unrealistically flattened images compared with simulations done on the
real data images.  However, the general detection parameters could
still be optimized with these more idealized simulations.  We found
that extremely low detection thresholds identified many random noise
peaks in the convolved image as real detections.  These preliminary
simulations also tested the minimum area parameter, {\tt
DETECT\_MINAREA},  and the kernel size of the convolution filter.  With
a large {\tt DETECT\_MINAREA} of 300 pixels we detected objects only
along a narrow central surface brightness regime, and lost a
significant number of small objects.  Using a {\tt DETECT\_MINAREA}  of
100 pixels, we recovered almost all input objects in the parameter
space tested.  Changing the kernel size from a 5\arcsec\ exponential to
a 10\arcsec\  exponential did not change the results dramatically,
except insofar as the smaller kernel pushed the limiting central
surface brightness a little fainter; thus, we adopted the smaller
filter size.

Ultimately, we expanded the simulations, adding the objects to the real
data frame for Field 7. The image has had the brightest galaxies
modeled and subtracted, and the halos of light surrounding the
brightest stars modeled and subtracted according to the procedure
described in
\addtocounter{section}{-1}\S\arabic{section}.2.\addtocounter{section}{+1}
Distributing galaxies in the real image probes not just the efficiency
of the detection parameters, but also loss of chip area due to very
bright objects and detection difficulties due to background
inhomogeneities.  We find that the latter difficulties with the
background dominate the detection process in the optimized method.

With real data images we further tuned our optimized search  method to
avoid possible false detections.  False detections can arise in the
masked and convolved image from peaks in the noise and from
flat-fielding artifacts ({\em e.g.} edges from the individual MOSAIC
CCDs).  Follow-up imaging of candidates and use of detectors with
better cosmetics than the MOSAIC engineering-grade chips will
ultimately eliminate almost all of these false detections.  In the
meantime, we can characterize the expected contamination from false
detections, and tune our detection threshold to exclude them, using a
negative version the current MOSAIC image.  We first inverted the
masked and background-subtracted version of the image, and then added
the background back in.  The resulting image is then convolved with the
same kernel, and the optimized detection method applied.  By inverting
the image, we insure that the only possible detections are artifacts
and not real objects.  False detections from noise should happen
equally often for the positive as for the negative image.  Similarly,
the MOSAIC CCD edge artifacts are the result of very slight
illumination mismatches --- both positive and negative --- between the
dithered images, so they create both positive and negative sharp edges
in the convolved image.  Other, more bizarre flat-fielding errors
create structure that is also typically symmetric around the background
level. Thus,  on average, the inverted image should result in the same
number of false detections from these types of contamination sources.
For Field 7, we set our detection threshold for the optimized method to
5.5-$\sigma_{sky}$, at which level we find zero possible false
detections that cannot easily be identified with an obvious flaw.

The total number of objects simulated in a
parameter bin, $n_{add}(R_T, \mu_0)$, is determined by the criterion that
the error in the final counts $N(R_T,\mu_0)$ contributed  
from the uncertainty in $f$\  is small compared to the error contributed by 
Poisson variance in the observed counts, $n_{obs}(R_T, \mu_0)$.  We adopt the error model for
deriving stellar luminosity functions corrected by artificial star
experiments of \citet{bol89}.  The uncertainty in a bin is given by  
\begin{eqnarray}
\sigma_N^2 & = & \frac{n_{obs}}{f^2} + \frac{(1 - f) n_{obs}^2}{n_{add} f^3}. 
\end{eqnarray}
With this model our criterion becomes
\begin{eqnarray}
\frac{(1 - f) n_{obs}}{n_{add} f} & \ll & 1,
\end{eqnarray}
and in practice we require
\begin{eqnarray}
n_{add} & \geq & 100 \times \frac{(1 - f) n_{obs}}{ f}.
\end{eqnarray}

We add a preliminary minimum number of 100 galaxies for
every bin and then run the experiments adding galaxies from
appropriate bins until Equation 4 is satisfied.

\subsection{Selection Function}
The explored parameter space presented in this paper, $17 \leq R_T \leq
22$, $22 \leq \mu_0 \leq 27$, is a subset of the final simulation
results to be presented in a subsequent paper, but encompasses the 
transition region between our traditional vs. optimized methods for the lowest
luminosity dwarfs.  The completeness fraction from these simulations
is plotted in greyscale in Figure \addtocounter{nfig}{1}\arabic{nfig} as a function of $R_T$\ and
$\mu_0$, where the greyscale levels correspond to 10\% (lightest),
30\%, 50\%, 70\%, and 90\% (darkest) completeness.  The first two
panels, Figures \arabic{nfig}a 
and \arabic{nfig}b, show $f$ as determined independently by the two
detection methods, traditional and optimized, respectively. The fact that the 
optimized method complements the traditional method can be seen in these contours,
where the optimized method pushes the 
completeness envelope to fainter surface brightness limits.  The
composite completeness factor combining both methods is represented in
Figure \arabic{nfig}c.    
LG galaxies if  seen at a distance of  10 Mpc are plotted for
comparison as filled circles, 
showing that at the $\sim$90\% completeness level we can detect dwarfs
comparable to Antlia and Sculptor and at the  $\sim$50\% completeness level we
can detect dwarfs similar to Tucana and Leo II.  The complexity of
the selection function shows  
that the completeness of our survey cannot be characterized by just a
magnitude limit or a surface brightness limit, but requires combining
both effects.  We can say that the faintest
central surface brightness we reach at the $\geq 50\%$ level is around
25.7 \magsqa, and the faintest magnitude is around $R_T \simeq 20$.  
In measuring profiles of our real and artificial galaxies, we 
robustly detect our objects to a limiting surface brightness isophote
of 26.7 mag/$\prime\prime^2$.

\subsection{Parameter Recovery}
The simulated galaxy experiments can also
give a measure of how robustly we 
measure  structural parameters.  For each object
recovered, the radial profile is non-interactively measured during the
simulation using the IRAF task 
{\tt ELLIPSE}.  We fit an exponential profile and determine the
extrapolated central surface brightness, 
$\mu_0$, and the fitted exponential scale length, $\alpha$.
We do not test our ability to 
discriminate between different profile shapes here.

In 
\addtocounter{nfig}{+1}
Figures \arabic{nfig}--\addtocounter{nfig}{+3}\arabic{nfig}, we
demonstrate  
our ability to recover
the input parameters central surface brightness, total magnitude, and
exponential scale length.  Our parameter recovery accuracy
 depends strongly upon the position of the galaxy in 
$R_T\  vs.\  \mu_0$ parameter space, and as expected, significantly
decreases as we approach our detection limits in either axis.   
\addtocounter{nfig}{-3}Figure \arabic{nfig} shows the difference
between input and output $\mu_0$ plotted as a
function of input total magnitude, with
the left panels showing all the artificial galaxies recovered by the
traditional method, and 
the right panels those recovered only by the optimized method.
In the vertical direction, we divide the artificial galaxies by input
central surface brightness: $22 \leq \mu_0^
{\rm in} < 23$, $23
\leq \mu_0^{\rm in} < 24$ and $24 \leq \mu_0^{\rm in} \leq 25$ from top to
bottom.  We can see our accuracy in parameter recovery decrease as the
detection efficiency starts to fall, near $R_T \simeq 20$.
The limits on $n_{add}$ imposed
by Equation \arabic{equation} can be seen, for example, in panels a, b,
and e, where there is a dramatic increase in the number of simulated
galaxies for certain bins.
 The complementary nature of the
two detection methods is also implicit in this figure and can be
compared to Figure
\addtocounter{nfig}{-1}\arabic{nfig}\addtocounter{nfig}{+1}.  Panel a 
shows the traditional method peaking in efficiency for 
$\mu_0^{\rm in} \sim 22.5, 17 \leq R_T \leq 19.5$, and panel f
shows the optimized method peaking for 
$\mu_0^{\rm in} \simeq 24.5, R_T \simeq 17.5$.  We see in the right-hand panels 
that the lowest-surface-brightness galaxies detected only
through the optimized method have a large scatter in $\Delta \mu_0$.
These objects are more difficult to fit
due to their very low S/N
which follows from their very low surface
brightnesses, especially as we approach our detection limits ({\em
e.g.} panel e). 

In \addtocounter{nfig}{+1}Figures \arabic{nfig} \&
\addtocounter{nfig}{+1}\arabic{nfig}, we show the magnitude recovery
from the simulations, comparing the input magnitude to 
the SExtractor-measured magnitude in 
\addtocounter{nfig}{-1}Figure \arabic{nfig}, and to 
the total magnitude calculated from the fitted parameters
in Figure 
\addtocounter{nfig}{+1}\arabic{nfig}.  In each figure 
we show the difference in input and
output total 
magnitude, $\Delta R_T = R_T^{\rm out} - R_T^{\rm
in}$, as a function of $R_T^{\rm in}$, with the same panel divisions
as the previous figure.
First, we recover the expected offset in SExtractor {\tt MAG\_BEST} in the
left-hand panels of Figure \addtocounter{nfig}{-1}\arabic{nfig}.  This
offset results from SExtractor's use of an aperture correction to approximate
the total magnitude which is known to systematically miss some light
at the $\sim 0.06$ magnitude level \citep{ber96}. Second, we see in
the right-hand panels of Figure \arabic{nfig}, especially in panel d,
evidence for the inability of SExtractor to de-blend objects properly when
run in double-image mode, as discussed previously. We see the faintest
galaxies tend 
to be measured brighter than they were input, as bright, neighboring
objects that fall within the photometric aperture are mistakenly being
included in the measurement.  This would also introduce larger
deviations from the true magnitude for fainter objects, and so we see
the scatter increase in panels e and f. In comparing the results from
the traditional method in Figures
\arabic{nfig} and
\addtocounter{nfig}{1}\arabic{nfig}, we see
that SExtractor {\tt MAG\_BEST} does a slightly better job of
measuring the magnitude than the total magnitude calculated from the
profile fits, which is why we adopt {\tt MAG\_BEST} for these
detections.  The larger scatter in $\Delta R_T$ seen in the left-hand
panels of Figure \arabic{nfig} are primarily due to difficulties in
measuring the exponential scale length in the radial profiles, as can
be seen in Figure
\addtocounter{nfig}{1}\arabic{nfig}\addtocounter{nfig}{-1}.     
The considerable scatter in the right-hand panels of Figure
\arabic{nfig} shows that, even inputting artificial exponential
profiles, the noise at these low surface brightnesses makes
automated profile fitting a difficult endeavor. 

In Figure \addtocounter{nfig}{1}\arabic{nfig} we demonstrate our
ability to recover the exponential scale length ($\alpha$) of the 
simulated galaxies with our profile fitting.Here we show the difference in input
{\em vs.} output exponential scale length, 
$\Delta \alpha = \alpha^{\rm out} - \alpha^{\rm in}$, as a function of
$R_T^{\rm in}$.  Panels a -- f are divided by 
central surface brightness in the same way as the previous figure.  
While the extrapolated
central surface brightnesses for these galaxies are well recovered, as
seen in Figure
\addtocounter{nfig}{-3}\arabic{nfig}\addtocounter{nfig}{+3}, we find
that the 
slopes of the profile fits, which yield the exponential scale lengths,
are much less well determined.  An unsuccessful profile
fit tends to approach the sky level, as background subtraction can
be difficult for a low S/N object.  On average this yields flatter
fits for more difficult objects, and thus larger
$\alpha^{\rm out}$.  We see this in our simulations, where the scatter
in $\Delta \alpha$ always tends towards $\alpha^{\rm out} > \alpha^{\rm
in}$.  Furthermore, the errors in $\alpha^{\rm out}$ seem to dominate the 
calculated total magnitude at all magnitudes and central surface
brightnesses, as the scatter in Figure \arabic{nfig} to preferentially
larger $\alpha^{\rm out}$ roughly corresponds to scatter in Figure
\addtocounter{nfig}{-1}\arabic{nfig}\addtocounter{nfig}{1} to preferentially brighter
$R_T^{\rm out}$ ({\em i.e.} $\Delta R_T \lesssim 0$).
Finally, we see again that for the majority
of the objects detected only through the optimized method (which are
typically of the lowest surface brightness or lowest S/N),
$\alpha$ is very difficult to determine.  

Some of our difficulties in
measuring the profiles of the faintest and lowest-surface-brightness
galaxies are mitigated   for our real candidate
detections by interactive profile
fitting and deeper follow-up imaging in the future.  We note that most photometric 
studies of galaxies with similar properties as our artificial galaxies
are likely affected to similar levels.

\section{Preliminary Results}
To demonstrate the success of our method, we present a preliminary
sample of objects detected in Leo
I.  First, we show the measured properties of objects 
found over the one-square-degree image of Field
7, and discuss how these properties can illuminate these objects'
group  membership.  Second, we 
examine several objects from the group in more detail, presenting 
examples of the types of supplementary observations we will also 
use to constrain membership and
contamination. 
Among these objects we include a newly discovered dSph in Leo I 
which has comparable
properties to a dSph like Sculptor ($M_R = -11.11, ~\mu_0 = 23.39~R$~ mag/$\prime \prime ^2$).
We also show exciting examples of how we may use the classic 
test of resolving our dwarf galaxies into stars to establish their
membership as done by \citet{baa44} to 
confirm M32 and NGC 205 as companions to M31, as well as using surface
brightness fluctuations (SBF) for relative distance determination.
Lastly, we show a 
preview of the contribution these detections make to the faint end of
the GLF.

\subsection{Field 7 Objects}

As mentioned previously, our object search process and membership
determination are guided by our knowledge of objects found in the
LG.  We know that most dwarfs at a
distance of 10 Mpc will subtend a large angle on the 
sky ($>20\arcsec$), plus appear distinctly diffuse
(LSB) with a 
roughly flat radial profile.  This was shown visually in
Figure 3. In addition, the more luminous  late-type
galaxies at 10 Mpc will have 
resolved structure such as spiral arms, patchiness and HII regions.
Early-type galaxies, the LG example being M32, have extremely bright
central surface brightnesses and steep radial profiles which
fall off quickly in angular size with increasing
distance. 
These morphological criteria, covering the most common morphological
types we find locally, allow for elementary membership assignment which
distinguishes group members from very distant background galaxies 
\citep[{\em c.f.} ][]{san84}.   Although these criteria have been very
successfuly employed, for instance, for the Virgo Cluster Catalog
\citep{dri96}, this morphological classification
can begin to break down for more nearby background galaxies, as well
as for more 
unusual types of galaxies not found in the LG. Such types of galaxies
are dwarf S0 
galaxies \citep{san84}, BCGs \citep{thu83} and their subset, the
so-called HII galaxies \citep{tel97}, and large, very
low-surface-brightness disks \citep{imp88}.  
In these less certain cases, we can begin to discriminate non-group-member
galaxies by further characterizing our sample 
using their measured properties such as total magnitude, central
surface brightness, and radial profile shape.

We present the properties of our detections from the entire
one-square-degree image of 
Field 7 
in \addtocounter{nfig}{+1}Figure \arabic{nfig}, plotted in the
same way as 
\addtocounter{nfig}{-6}Figure \arabic{nfig}. The detections in this
figure include 
contamination from background non-group-members, which appear most 
prominently 
near $R_T \sim 19.5, \mu_0 \sim 22$. 
The crosses show all detections from the traditional method 
after star/galaxy separation (as in 
Figure \arabic{nfig}\addtocounter{nfig}{+6}). The plotted  parameters
for these objects (the crosses) 
are the total magnitude as measured by SExtractor and the central
surface brightness as extrapolated from the radial profile fit. 
Objects detected through the optimized method are shown as
stars.  The plotted parameters for these detections are both calculated
from the profile fit.  Objects detected by  both methods are connected
by a line.
For comparison we show again LG dwarfs scaled to a distance of
10 Mpc, which are marked by filled circles. 

Again we see the effectiveness of this plot as the smaller, more
compact objects  
that are mostly distant background galaxies separate out from the
expected locus of LG-like dwarfs. 
This also follows from the radial profiles, and thus morphologies, we
expect from known galaxy 
types, which can be qualitatively seen in Figure 1. For example,
early-type, de-Vaucouleurs-profile galaxies do not populate this LSB regime of 
parameter
space shown in Figure \arabic{nfig} unless they are very distant,
cosmologically dimmed objects, and 
even then their small angular sizes would make them virtually undetectable
within our limits. 

The galaxies circled in Figure \arabic{nfig} are described in more
detail in 
\addtocounter{nfig}{+1}Figures \arabic{nfig} --
\addtocounter{nfig}{+3}\arabic{nfig}, where we show a 
2\arcmin$\times$2\arcmin\ postage stamp  of the MOSAIC $R$-band image for
each example and its radial profile measured with {\tt ELLIPSE}. 
\addtocounter{nfig}{-3}
In Figure \arabic{nfig}a we
show one of the larger, higher-SB galaxies in Field 7.  This is a
previously catalogued object, CGCG 065-091 \citep{zwicky}, which 
has a measured velocity listed in 
NED\footnote {NASA/IPAC Extragalactic Database} of 7922 \kms\
\citep{fou92}. Note that it shows a sign of spiral structure, but
without the  
well-defined spiral arms of NGC 3351.  Neither does
it have the expected size and magnitude of a Sbc galaxy at 10 Mpc.
The radial profile of CGCG 065-091 is shown in Figure \arabic{nfig}b,
where the disk and bulge components can be seen.
Based upon its morphology and properties, this
galaxy is likely to be in the background of the Leo I group, and this
fact is corroborated by its velocity.
The previously uncatalogued galaxy shown in
Figure \addtocounter{nfig}{+1}\arabic{nfig}a is a likely non-member
from the other extreme of parameter space,
having a 
small isophotal size as indicated by the detection aperture, and
having photometric properties ($R_T = 19.66, \mu_0 = 22.04$) in Figure
\addtocounter{nfig}{-2}\arabic{nfig} outside those expected for 
LG-type dwarfs.  
\addtocounter{nfig}{+2} The profile in Figure \arabic{nfig}b is mostly
dominated by seeing, due to its small angular size at our isophotal
limits, but suggests an 
exponential profile.  The profile fit parameters, however, are
inconsistent with those expected for LG-type dwarfs ($\alpha^{\rm fit}$ =
1\farcs41, \mucfit = 21.95).

In \addtocounter{nfig}{+1}Figures \arabic{nfig} and 
\addtocounter{nfig}{+1}\arabic{nfig} we show two objects which are
each detected by both the traditional and the optimized method.
 In each figure, we show in panel (a) the
original MOSAIC image with the 
detection aperture from the traditional method indicated, in 
panel (b) the masked image with the detection aperture for the  
optimized method indicated and in panel (c) the
convolved image with the optimized detection aperture again indicated.
A comparison of the aperture sizes in panels (a) and (c) directly
contrasts the two detection 
methods: the detection threshold of the traditional method in (a) only
detects the bright center of the galaxy, whereas the optimized method
in (c) also detects an extended, low-surface-brightness component.
The extremely low surface brightness and extended sizes of these
objects suggest they are dSphs in Leo I, and the object in Figure
\addtocounter{nfig}{-1}\arabic{nfig} might be a nucleated
dwarf.

We plot the radial profiles for
these two objects in panel d of Figures \arabic{nfig} and
\addtocounter{nfig}{+1}\arabic{nfig}.  The profiles extend to $\mu(r)
\simeq 26~R$~mag/$\prime \prime ^2$ out to $\sim 1.6\alpha^{\rm fit}$
for Figure
\addtocounter{nfig}{-1}\arabic{nfig}d\addtocounter{nfig}{+1}, and
somewhat further in Figure \arabic{nfig}d, $\mu(r)\simeq 28 R$
mag/$\prime \prime ^2$ out to $\sim 3.4\alpha^{\rm fit}$.  We fit both
profiles with an exponential component, adopting \mucfit and
$R_T^{\rm fit}$.  In Figure
\addtocounter{nfig}{-1}\arabic{nfig}d\addtocounter{nfig}{+1}, where we
see the suggestion of a nucleus, we fit only the exponential
component.  If the nucleus is real, it should contribute only a small
amount to the total flux of the object ($\sim15\%$ for the faintest
galaxies), as has been seen for nucleated 
dwarfs found in the Virgo cluster \citep{san84,imp88}.  In addition,
excluding the nucleus from the fit makes the underlying dwarf
characteristics more directly comparable to dwarf counterparts in the
LG (although, the nature of the nucleus and its relationship to the
underlying stellar population is still not well known).
The final parameters of these galaxies support their classification as
dwarf spheroidal members of Leo I.

\subsection{Supplementary Observations}

Although we have some confidence in the identification
of Leo-group dwarf galaxy members on the basis of 
morphology, total magnitude and central surface brightness
({\em i.e.} the location of candidates in Figure
\addtocounter{nfig}{-4}\arabic{nfig}\addtocounter{nfig}{+4}), we are  
planning and engaged in  follow-up observations.
To measure a completely reliable GLF, we would like to verify
membership for every Leo member candidate. Particularly for the
low-surface-brightness candidates, this is an ambitious
undertaking. One important side benefit of the follow-up observations
is that when our final list of Leo Group galaxies is assembled,
we will have a significant database of the structural, photometric,
and in some cases, spectral properties of the ensemble. 
We have three on-going programs described in the following subsections.

\subsubsection{Multi-color imaging}

Follow-up imaging of our candidates can provide a fuller
characterization of very low-luminosity galaxies in an environment
similar to, but distinct from, the LG.   In addition,
the lowest-surface-brightness candidates, which are primarily detected 
through our filtering search technique, can be subject to
contamination by some false detections and high-$z$ galaxies and
galaxy clusters. 
Because of the difficulty in flat-fielding the MOSAIC images
and the difficulty in completely removing chip flaws and the
intra-CCD dead regions in the final stacked image, some of
our candidates in the MOSAIC fields are going to be spurious.
Imaging each candidate with another telescope will quickly
allow us to cull such detections from our list.  Our filtering
technique also makes us sensitive to background LSB disk galaxies \citep[we
expect $\sim4$/deg$^2$,][]{dal97},
cosmologically-dimmed $z > 0.4$
galaxies and $z > 0.6$ galaxy clusters ($\lesssim1$/deg$^2$;
A. H. Gonzalez 2000, private communication).
These candidates are difficult to reach
spectroscopically; however, with follow-up
imaging in $B$ and $I$, the three colors (we have
$R$ from the 0.9m+MOSAIC imaging) can
flag higher-$z$ galaxy and cluster contamination due to their very red
colors.
In Figure \addtocounter{nfig}{+1}\arabic{nfig}, adapted from
\citet{fuk95}, we show the distribution of $B-R$ for various
morphological types  as they change with redshift (generally increasing to the
right).  The circled
values are for $z = 0$, squares are $z= 0.2$, triangles are $z=0.5$
and hexagons are $z=0.8$.  Measured values for LG galaxies
today fall  along the $z=0$ locus.
Note that all objects with $B - R > 1.7$, according to these models, have
$z \ge 0.2$.   Multiple
colors can provide a broader baseline for 
such  
discrimination, although such color classification must also be
approached conservatively, as such models do not account for unusual
objects, such as the very red LSB galaxies recently discovered by
\citet{one00}.

\subsubsection{Spectral observations}

It is possible to obtain spectra suitable for measuring
radial velocities for a large fraction of the candidates.
We have already devoted two Keck nights and two Lick 3m nights
to this effort with many more required  to complete the follow-up of
the Leo Group candidates. Figure \addtocounter{nfig}{+1}\arabic{nfig}
shows an example candidate (2$^\prime\times2^\prime$ postage stamp
from our MOSAIC image) 
and its spectrum. In this case the spectrum was obtained with the Lick 3m+KAST
spectrometer in a $3\times$900s exposure, and the velocity calculated
via the cross-correlation with a velocity template. The velocity in this case,
6525 \kms,
showed this object to be background to the group. Figure 
\addtocounter{nfig}{+1}\arabic{nfig}
shows a more challenging example, which is an object previously
catalogued by \citet{fs90} corresponding to LEO \#13
in their catalog list.  This spectrum required 
1.5 hours integration with Keck II$+$LRIS and 
cross-correlating with a velocity template to extract a velocity. For
this object, one of the  
best candidate dSph galaxies, we measure the velocity of 886 \kms,
indicating
this is a bona-fide Leo Group dwarf.

\subsubsection{Red Giant Branch Tip and Surface brightness fluctuations}

At a distance of $\sim 10$ Mpc ($m-M \sim 30$) it is possible to use two 
additional photometry-based
methods to test for group membership for our candidate Leo dwarfs.
The brightest of the first-ascent red-giant-branch stars, at $M_I\sim -4$, will
be measurable at $m_I\sim 26$ mag in Hubble Space Telescope images or
with large ground-based telescopes and good seeing. \citet{sak97}
used a modest amount of HST observing time to demonstrate this for 
the Leo Group giant elliptical NGC 3379. Their observations were made
in an outer region of the galaxy at a surface brightness ($\mu_B\sim 27.1,
\mu_I\sim 25.2$) which is very similar to the central surface brightness of our
most diffuse candidates. The tip of the red-giant branch (TRGB) was easily detected.

Measurement of surface brightness fluctuations has proven to be a
powerful method for estimating distances to galaxies to beyond
the distance of the Virgo cluster and also as a means of inferring
the stellar populations in galaxies. 
NGC 3379 in the Leo I Group has also been used to 
to demonstrate this technique \citep{ton88}. \citet{bot91} showed that
low-surface-brightness dwarf galaxies in the Fornax cluster are particularly
well-suited for use of this technique and more recently it has been
applied specifically to low-surface-brightness dwarfs 
\citep[{\em e.g.} ][]{one99,jer00}.
This method has the potential
for allowing membership to be determined for the most challenging of the
Leo LSB dwarf candidates --- those with the lowest central surface brightness. 
Although there is potential for learning about the stellar
populations in any Leo Group members, we will initially be performing
a binary test for membership. The candidate Leo I galaxies will generally
either be near 10 Mpc or at a much larger distance in the background.
The brightest giants will be resolved or
not; the LSB galaxies will show measureable fluctuation or not.

In March 2000, we imaged several of the dSph candidates with the new
imaging spectrograph `ESI' on the Keck II telescope. With these data we
can show the promise of the TRGB and SBF techniques for our specific
application. For three of the galaxies, imaged under excellent
conditions with point sources showing $<0\farcs5$ FWHM, the galaxies
just barely resolved into stars in only a 10min exposure in the $I$ band.
Figure \addtocounter{nfig}{+1}\arabic{nfig} shows an example of one of
our `resolved' candidates (the same object as in Figure
\addtocounter{nfig}{-1}\arabic{nfig}\addtocounter{nfig}{+1}).
Such excellent seeing conditions are not to be counted on
for ground-based imaging and we do not expect to be able to
survey a large number of dwarf candidates with this technique from
the ground.  With HST however, the TRGB stars in Leo Group LSB dwarfs
can be resolved in approximately 20 min observing time.

It is possible to measure SBF for Leo I dwarfs with ground-based imaging.
Figure \addtocounter{nfig}{+1}\arabic{nfig} shows the fluctuation
power spectrum for the galaxy from the previous example. 
In Figure \arabic{nfig}, we compare the power spectrum
from a section of the galaxy with no discernable brightness
gradient (filled boxes) to the power spectrum of a
star (stars) and to a blank section of the sky (triangles).   
Note that at large $k$, you see the constant signature of the white
noise spectrum, whereas at small $k$, you see the PSF-like signature of
both the star and the underlying unresolved stars in
the galaxy image.  This contrasts with the spectrum of the sky,
where there is only a very weak PSF contribution from unresolved
background galaxies. We have also estimated the amplitude of
fluctuations via the method described in \citet{bot91}.
The variance in counts was measured in multiple 100-pixel boxes
placed in `sky' regions of the frame and in six positions within the
galaxy after point sources were removed and a median-smoothed 
version of the image was subtracted out to remove the large-scale
gradients in the galaxy profile. The excess fluctuation signal above
that expected in the sky plus galaxy is $\sim 9$\% of the galaxy
signal. We are unambigously measuring the pixel-to-pixel variations
due to the small number of average-luminosity giants over the face of the
galaxy. Rather than attempting to derive an absolute calibration for
distance based on our SBF measurements in the Leo I candidates, we
will be able to make a differential distance estimate comparing to
SBF measured at different surface brightness levels in NGC 3379 (and
possibly correcting for stellar population differences based on
color).

\subsection{Preliminary GLF}

\addtocounter{nfig}{+1}
With our preliminary membership assignment, we can begin to measure
the GLF.  In Figure \arabic{nfig} we present a histogram of absolute
$R$-band magnitudes \citep[assuming $m-M = 30$ based on 
cepheid measurements
in NGC 3351;][]{gra96}
for objects found in Field 7 that are likely members.  While this is
not a robust measurement of the galaxy luminosity density for the
group (we reserve this analysis to a later paper with the full group
sample), we note that even with a preliminary membership
classification, we probe fainter magnitudes than have been studied in
the Leo I group to date. The nominal 100\% completeness limit of
\citet{fs90} is $M_{B,lim} = -14.2$ (Ferguson \& Sandage values adjusted
for $m-M = 30$), which for 
$B - R $ = 1.3 corresponds to $M_{R,lim} = -15.5$.  Their
photographic survey, which covers a comparable area on the sky, 
used an isophotal diameter limit of 17\arcsec\ at 
$\mu_B = 27$ \magsqa, and beyond this limit they applied an
incompleteness correction 
based upon the expected range of the $\mu_e - B_T$ relation as measured in
Fornax 
\citep{fs88}.  They find this correction to be small out to 
$M_B \simeq -11.2$ ($M_R \simeq -12.5$), although their completeness description does not
account for objects of lower surface brightness for a given magnitude
than predicted by the $\mu_e - B_T$ relation like those recently
discovered in Fornax \citep{kam00}.  This would suggest a larger
incompleteness fraction at lower magnitudes, as well as perhaps an even
steeper faint-end slope than the $\alpha = -1.4$ predicted in
\citet{fs91}.  One of the real advantages of our simulation method is
that we  explore these possible surface brightness selection
effects.  We see from our simulations, and from our candidate lists
that, even at faint surface 
brightnesses, we can detect objects to $M_R > -11$.

\section{Discussion}

Even without full membership determination, we have
demonstrated the success of our method in discovering 
dwarf spheroidal galaxies in the Leo I group.  With the advent of
MOSAIC imaging, we combine a very wide-field survey with the
advantages of CCD photometry and can now explore nearby galaxy groups
to the same sensitivities as were previously available only for
smaller, and so more distant, galaxy groups and clusters.
At the 70\%
completeness limit for this field, we extend
our survey at least 2 magnitudes beyond the
photographic work done by \citet{fs91} for $\mu_0 \lesssim 24.5$, 
discovering 
dwarf spheroidals like  that shown in Figure 16.  We also find an
example (Figure 15)  of the large, LSB dwarfs first discovered in Virgo
\citep{imp88} --- objects which are apparently absent from the Local
Group. This object corresponds to LEO \#5 in the
\citet{fs90} catalog. 

One possible limitation of our sample selection is that our
methods 
are optimized
to search for the typical objects we would expect in the Local Group. 
This can lead to some prejudices.   For instance, 
very compact, faint objects we classify as background, assuming the faintest
galaxies will be diffuse dwarf spheroidals.  Overall, this is justified
since the only type of compact dwarf seen the local Universe 
is M32-like, and we detect no objects with such high central surface
brightnesses that SExtractor classifies as galaxies (distant
objects with these bright $\mu_0$ are so compact they are classified
as stars by SExtractor).  Furthermore, all
similarly compact objects detected in the 
Leo I group by \citet{fs90}, which were classified as possible M32-like
candidates but also as non-members, were later
spectroscopically confirmed as background galaxies \citep{zie98}.  
However, unusually compact objects not seen in the
LG, or BCDs, might be lost from our sample in this way. Other 
objects not seen in the LG are the large, LSB dwarfs seen in 
Virgo \citep{imp88}, Fornax \citep{bot91} and M81 \citep{cal98}.  In
Fornax a population of  extreme LSB dwarfs 
was detected that could rival the brighter galaxies in integrated
luminosity and mass \citep{kam00}.  Although we don't preferentially
search for this type of galaxy, we find from our simulations 
that the sensitivity of our search 
method peaks for bright ({\em i.e.} large) LSB dwarf galaxies like these.  
So, if such objects exist in large numbers in Leo I (we find only one in
Field 7), we will detect them.

Our understanding of the LG also leads
us to expect  
dwarfs to cluster as satellites around brighter, companion galaxies.
This correlation is seen in some other low density 
environments \citep{vad91,lov97}, although there are  hints that
the dwarf-giant correlation is lowest in poor group environments,
where dwarfs might even be `free-floating' \citep{vad92}.  To explore
this, we have tried to 
recover as much  
area as possible surrounding the large galaxies in our sample, but for
late-type, $L^*$ galaxies like NGC 3351 in Field7, there would be significant
confusion of diffuse
dwarfs with any residual structure from the profile subtraction, such
as HII regions and spiral arms. Thus, we ignore any detection within a
projected distance of $\sim4$\arcmin\ (11.5 kpc) around NGC 3351.  Despite this
limitation, however, we would still expect to detect a significant
fraction of a satellite dwarf population.  The characteristic distance
of a Milky Way dwarf satellite is at a projected distance of 141 kpc
from the Galaxy 
\citep{kar96}, 
which would correspond to $\sim48$\arcmin\ at 10 Mpc, and the
corresponding distance for M31 satellites is 85 kpc \citep{kar96}, or
$\sim30$\arcmin\ at 10 Mpc.  As we do not see the Milky Way satellites
in projection, it is difficult to quantify how many would fall too
close to the Galaxy to be detected. In the case of M31 satellites, two
objects 
would fall within 4\arcmin\ of the main galaxy, but these are M32 and NGC
205 which cause little concern for confusion.

The primary limiting factor for these data is the nature of the
engineering-grade CCDs used in the MOSAIC camera for the majority of
our Leo I images.  Even modest brightness enhancements due to lack of
flatness of the images can be enhanced
through our filtering process, forcing us to raise our detection
threshold to exclude possible spurious detections.  With flatter images, we
can lower our threshold and push our limits to fainter surface
brightnesses and magnitudes. MOSAIC images of our Fields 8 and 9 were
taken with the 
newer science-grade CCDs last Spring, and will provide this
comparison.  Our current limitations are well quantified, however,
with our simulations.

The galaxy simulations not only highlight selection effects in
detection, but also our ability to  measure photometric
parameters for these candidate galaxies.  As we have shown in \S4.3,
parameter recovery can be rather difficult for the lowest-luminosity
and lowest-surface-brightness objects.  Part of this difficulty is in using an
automated process, and for all real objects detected through the
optimized method, we opt to fit them by hand.  Full understanding of
these challenges to our photometry provides a robust and direct
measurement of our errors.

The discovery of dwarf spheroidal galaxies outside the Local Group is a
burgeoning pursuit that has begun to yield tantalizing results in a
number of environments \citep[{\em e.g.} Virgo and
Dorado clusters;][]{phi98,car01}.  With theoretical models predicting a steep
faint-end slope to the GLF \citep{col00} and detections of possible
bright dSph progenitors seen at $z > 0.2$ \citep[][]{guz98}, it is
critical to establish a census of these low-luminosity galaxies at
$z=0$.  As 
we clearly establish the selection parameters influencing our sample,
we can construct the only complete sample of dwarf galaxies to these
luminosities and surface brightnesses outside of the Local Group.
Furthermore, with such a sample we may explore the intrinsic nature of
these galaxies as an ensemble, such as their spatial distribution,
abundances, and stellar populations.


\acknowledgements
We thank the referee, G. Bothun, for useful comments and a careful
reading of the manuscript.  KF thanks A. Gonzalez for many helpful
discussions.  MB acknowledges support from the National Science
Foundation through the grant AST-9901256 which has funded most of this
work.  KF and AJM also acknowledge support from the California Space Grant 
Consortium through a fellowship and a mini-grant. CMO thanks the
Brazilian funding agency FAPESP for financial support.
This research has made use of the NASA/IPAC Extragalactic Database
(NED) which is operated by the Jet Propulsion Laboratory, California
Institute of Technology, under contract with the National Aeronautics
and Space Administration.

\clearpage

\figcaption[figure1.gif]{{\bf Upper:} The typical ranges of 
$R$-band central surface brightness vs. total absolute
magnitude  occupied by local spirals,
ellipticals, globular clusters, irregulars and dwarfs, 
adopted from Ferguson \& Bingelli 
(1994) Figure 3.  Plotted points are the members of the Local
Group, taken from Mateo (1998, assuming $B - R = 1.3$ for dwarfs
without a $R$ magnitude) 
and in the case of And V, And VI and
And VII taken from Caldwell (1999).  For the Milky Way and M31, the central
surface brightnesses are from Freeman's Law (Freeman 1970).  Local Group
galaxies are coded by 
morphological type: five-point stars are Spirals, open triangles are
Irrs, filled triangles are 
dIrrs, circled filled triangles are transition dSph/dIrr and filled
circles are dSphs.  The parameters of M32 and Malin 1 are plotted
individually to demonstrate their extremity.  {\bf Lower:} A blow-up of the 
region of parameter space that our Leo I survey will probe. Local
Group galaxies are now plotted with central surface brightness vs. apparent
$R$-magnitude as they would be seen at the distance of Leo I (10 Mpc).  
Symbols are the same as in the upper panel.  The solid lines represent
lines of constant angular size at an isophotal limit of $\mu =$
26.7mag/$\prime \prime^2$, assuming an exponential profile (Allen \&
Shu 1979).  Dashed lines are lines of constant exponential scale
length, for a purely exponential profile.
}

\figcaption[figure2.gif]{Our coverage of the Leo I group, overlaid on an image of
Leo I from the DSS.  We show our nine $1\arcdeg \times 1\arcdeg$
fields imaged in $R$ with the KPNO 0.9m$+$MOSAIC. For reference, NGC
3379 can be seen near the center of Field 4.  We use Field 7 in an
example of our search technique and first results.
}

\figcaption[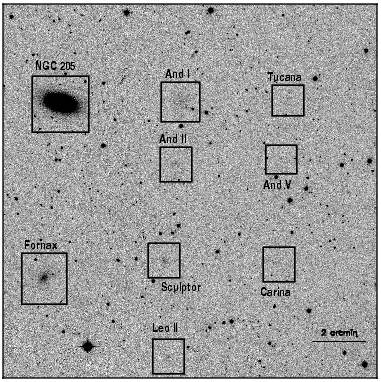]{A $15\arcmin \times 15\arcmin$  sub-image of
Field 7 with artificial galaxies added that mimic the characteristics
of Local Group dwarfs if seen at a distance of 10 Mpc.  Note that
Leo II, Tucana, And V, and Carina would all be difficult to detect by eye.
}

\figcaption[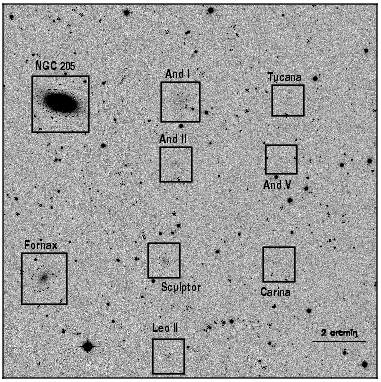]{The results of Step 1 of our detection method
on the image from Figure 3.  A SExtractor aperture is drawn around
every object detected that has been classified as a galaxy (see
text).  Most artificial dwarfs are detected, except for galaxies with
the characteristics of And II, Tucana,
And V and Carina, which have central surface brightnesses fainter than
the detection threshold.
}

\figcaption[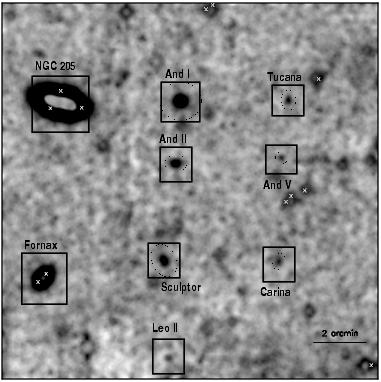]{The results of Step 2 of our detection method,
where we show the masked and convolved version of 
the image from Figure 3.  All SExtractor detections are indicated, where
obvious masking remnants from stars or bright galaxies are marked with
crosses and candidate detections are marked with SExtractor apertures.
Note that in this method, And II, Tucana, And V and Carina are now
robustly detected; however, we lose Leo II due to its center being
masked.  This demonstrates the necessity of using both methods
together. 
}

\figcaption[figure6.gif]{Total apparent magnitude vs. central surface
brightness for the detections from Figures 4 \& 5.  Detections from
the traditional method are shown as crosses and detections from the
optimized method as stars.  Local Group dwarfs as seen at 10 Mpc are
plotted as filled circles.  The inputted artificial galaxies from
Figures 4 \& 5 are labeled and connected by a line to their
measured parameters from both methods.  The solid and dashed lines are
the same as in Figure 1.
}

\figcaption[figure7a.gif,f7b.gif,f7c.gif]{{\bf a:} Detection
efficiency contours from our simulations of 
recovering input artificial dwarfs in Field 7.  These contours are for
objects recovered only by the traditional method.  The greyscale shows
recovery fractions of: 90\% (darkest), 70\%, 50\%, 30\%,
10\%(lightest).  Local Group galaxies if seen at 10 Mpc are again
plotted as filled circles, and the solid and dashed lines follow
Figure 1.
{\bf b:} Detection efficiency contours from the simulations,
showing the efficiency of detecting objects that are only
found with the optimized method.  Greyscale, symbols and lines are the
same as in Figure 7a.
{\bf c:} Detection efficiency from simulations for the combined
methods.  Greyscale, symbols and lines are the
same as in Figures 7a \& 7b.
}

\figcaption[figure8.gif]{Parameter recovery  as measured by the
simulations on Field 7, showing the difference in
output  and input central surface brightness ($\mu_0^{out}\ {\rm and}\ \mu_0^{in}$,
resp.) as a function of input 
total magnitude, $R_T^{in}$. The output central surface brightnesses are extrapolations
from exponential fits to the radial profiles.  The left panels
are all the objects detected through 
the traditional method, while the right panels are the object only
detected through the optimized method.  Objects are also divided by
input central surface brightness in the vertical direction.  
}

\figcaption[figure9.gif]{Recovery of total magnitude as measured by
simulations, where the output total magnitudes ($R_T^{out}$) for both
detection methods are {\tt MAG\_BEST} from SExtractor. 
The difference in output and input magnitude is plotted as a function
of input total magnitude ($R_T^{in}$).  The divisions by 
panel are 
the same as in  Figure 8.  
}

\figcaption[figure10.gif]{Recovery of total magnitude, as in Figure 9, except the
output total magnitudes plotted for both methods are now calculated
from the  exponential profile fits.  Panel divisions are the same as
in Figure 8.
}

\figcaption[figure11.gif]{Recovery of exponential scale length ($\alpha$) from simulations,
where output exponential scale length ($\alpha^{out}$) is measured from the exponential
profile fit of recovered objects.  Again we plot the difference in
output and input $\alpha$ as a function of input total magnitude.
Panel divisions are the same as in Figure 8.
}

\figcaption[figure12.gif]{Same as Figure 6, except here we plot all detections from
Field 7.  Objects detected by both methods are
connected with a line.  Example objects discussed further in \S 5 are
circled. 
}

\figcaption[figure13.gif]{Detected background spiral, CGCG 065-091.  {\bf a:} $2\arcmin
\times 2\arcmin$\ postage stamp  of the MOSAIC $R$-band image, with
detection aperture from SExtractor indicated.  {\bf b:} Radial profile
with errorbars as 
measured by {\tt ELLIPSE} in IRAF. Fitted exponential disk and
$R^{\frac{1}{4}}$-fit overlaid.
}


\figcaption[figure14.gif]{Detected distant background objects --- probably a distant
spiral.  {\bf a:} $2\arcmin \times 2\arcmin$\ MOSAIC $R$-band image,
as in Figure 13.  {\bf b:} Radial profile of object, as in Figure 13.
}

\figcaption{[figure15.gif]Candidate nucleated dwarf. {\bf a:} $2\arcmin
\times 2\arcmin$\ MOSAIC $R$-band image, showing the traditional
method detection aperture.  {\bf b:} The same image, masked of bright
objects, with the optimized method aperture overlaid. {\bf c:} The
same image as in b, convolved, with the optimized aperture overlaid.
{\bf d:} The measured radial light profile with exponential fit drawn
as a line. Fitted exponential scale length,
$\alpha^{\rm fit}$, and half-light radius, $r_{\rm eff}^{\rm fit}$,
are labeled.
}

\figcaption[figure16.gif]{Newly discovered dwarf spheroidal.  Panels same
as in Figure 15.
}

\figcaption[figure17.gif]{A diagram of model $B- R$ values 
from Fukugita et al. (1995), which shows the distribution of color for
each morphological type as it changes with redshift. The circled
values are for $z = 0$, squares are $z= 0.2$, triangles are $z=0.5$
and hexagons are $z=0.8$. Measured values for Local Group galaxies
today correspond to values  along the $z=0$ locus.
Note that all objects with $B - R > 1.7$ have
$z \ge 0.2$.
} 

\figcaption[figure18.gif]{Candidate galaxy found to be in the background of Leo I.
{\bf a:} 2$^\prime\times2^\prime$ image as in Figure 13a.  {\bf b:}
$3\times$900s spectrum taken with Lick 3m$+$KAST.  Velocity measured is
6525 \kms, which is far in the background of the group.
}

\figcaption[figure19.gif]{Candidate dwarf confirmed to be in Leo I.  {\bf a:}
2$^\prime\times2^\prime$ image from MOSAIC. Labeled parameters are
from profile fit.  {\bf b:} 1.5 hour spectrum with  Keck II$+$LRIS.
Measured velocity is 886 \kms, placing this dwarf well within the Leo
I group.
}

\figcaption[figure20.gif]{$I$-band image of Leo I candidate dwarf shown in previous
figure. This image was taken on  Keck
II$+$ESI with extraordinary seeing ($<0\farcs5$ FWHM), and quickly
resolved into the brightest stars.  Removing the point sources from
this image, we can measure the surface-brightness-fluctuation signal. 
}


\figcaption[figure21.gif]{The SBF power spectrum of the object in Figure 20.  The
filled boxes are the power spectrum measured from a patch of the galaxy, the
stars are the spectrum measured from a bright star in the image and the
triangles are the spectrum we measure from a randomly chosen empty
region of sky.  The galaxy spectrum (boxes) shows we are measuring a
PSF-like signal from 
unresolved stars in the galaxy which traces the PSF spectrum from the
star (stars) and far exceeds the very weak
PSF-signal from unresolved 
background galaxies in the empty sky image (triangles).
}

\figcaption[figure22.gif]{A histogram of detected objects in Field 7, applying a
preliminary morphological  membership classification.  While this is
not yet a robust measurement of the Leo I luminosity function, it
demonstrates the depth of our survey.
}


\begin{thebibliography}{}

\bibitem[Allen \& Shu(1979)]{all79}Allen, R. J.  \& Shu, F. H.  1979,
\apj, 277, 67

\bibitem[Armandroff, Jacoby, \& Davies(1999)]{arm99} Armandroff, T. E., Jacoby,
G. H., \&  Davies, J. E. 1999 \aj, 118, 1220

\bibitem[Baade(1944)]{baa44}Baade, W. 1944, \apj, 100, 137

\bibitem[Bernstein et al.(1995)]{ber95} Bernstein, G. M., Nichol, R. C.,
Tyson, J. A., Ulmer, M. P., \& Wittman, D. 1995, AJ, 110, 1507

\bibitem[Bertin \& Arnouts(1996)]{ber96} Bertin, E. \& Arnouts,
S. 1996, A\&AS, 117, 393

\bibitem[Binggeli \& Cameron(1991)]{bin91} Binggeli, B. \& Cameron,
L. M. 1991, \aap, 252, 27 

\bibitem[Bolte(1989)]{bol89}Bolte, M. 1989, \apj, 341, 168

\bibitem[Bothun et al.(1991)]{bot91}Bothun, G., Impey,
C., \& Malin, D. 1991, \apj, 376, 404

\bibitem[Caldwell(1999)]{cal99} Caldwell, N. 1999, \aj, 118, 1230

\bibitem[Caldwell et al.(1998)]{cal98}Caldwell, N., Armandroff, T. E.,
Da Costa, G. S., \& Seitzer, P. 1998, \aj, 115, 535

\bibitem[Caldwell \& Bothun(1987)]{cal87}Caldwell, N., \& Bothun,
G. D. 1987, \aj, 94, 1126

\bibitem[Carrasco et al.(2001)]{car01}Carrasco, E. R., Mendes de Oliveira,
C., Infante, L., \& Bolte, M. 2001,\aj, in press (astro-ph/0010076)

\bibitem[Christlein(2000)]{chr00}Christlein, D. 2000, \apj, in press
(astro-ph/0006450) 

\bibitem[Cole et al.(2000)]{col00}Cole, S., Lacey, C. G., Baugh,
C. M., \& Frenk, C. S. 2000, \mnras, 319, 168

\bibitem[Dalcanton et al.(1997)]{dal97}Dalcanton, J., Spergel, D. N.,
Gunn, J. E., Smith, M., \& Schneider, D. P. 1997, \aj, 114, 635

\bibitem[Dalcanton(1995)]{dal95}Dalcanton, J. J. 1995, PhD Thesis, Princeton
University 

\bibitem[de Propris \& Pritchet(1998)]{deP98}de Propris, R. \&
Pritchet, C. J. 1998, \aj, 116, 1118

\bibitem[Drinkwater et al.(1996)]{dri96}Drinkwater, M. J., Currie,
M. J., Young, C. K., Hardy, E., \& Yearsley, J. M. 1996, \mnras, 279, 595 

\bibitem[Faber \& Lin(1983)]{fab83} Faber, S. M. \& Lin, D. N. C.  1983,
\apj, 266, L17 

\bibitem[Ferguson \& Bingelli(1994)]{fer94}Ferguson, H. C. \& Bingelli,
B. 1994, \araa, 6, 67

\bibitem[Ferguson \& Sandage(1988)]{fs88} Ferguson, H. C. \& Sandage,
A. 1988, \aj, 96, 1520

\bibitem[Ferguson \& Sandage(1990)]{fs90} Ferguson, H. C. \& Sandage,
A. 1990, \aj, 100, 1

\bibitem[Ferguson \& Sandage(1991)]{fs91} Ferguson, H. C. \& Sandage,
A. 1991, \aj, 101, 765

\bibitem[Fouqu\'e et al.(1992)]{fou92}Fouque, P., Durand, N.,
Bottinelli, L., Gouguenheim, L., \& Paturel, G. 1992, Catalogue of Optical
Radial Velocities, Vol. 1 (Paris: Observatoires de Lyon et Paris-Meudon)

\bibitem[Freeman(1970)]{fre70}Freeman, K. 1970, \apj, 160, 811

\bibitem[Fukugita, Shimasaku, \& Ichikawa(1995)]{fuk95} Fukugita, M.,
Shimasaku, K., \& Ichikawa, T. 1995, PASP, 107, 945 

\bibitem[Garcia(1993)]{gar93}Garcia, A. M. 1993, \aaps, 100, 47

\bibitem[Graham, et al.(1996)]{gra96}Graham et al. 1996,
\apj, 477, 535 

\bibitem[Guzman et al.(1998)]{guz98}Guzman, R., Jangren, A., Koo,
David C., Bershady, M. A., \& Simard, L. 1998, \apj, 495, L13

\bibitem[Hradecky et al.(2000)]{hra00}Hradecky, V., Jones, C.,
Donnelly, R. H., Djorgovski, S. G., Gal, R. R., \& Odewahn, S. C. 2000, in
press (astro-ph/0006397)

\bibitem[Impey \& Bothun(1997)]{imp97}Impey, C. \& Bothun,
G. 1997, \araa, 35, 267

\bibitem[Impey et al.(1988)]{imp88}Impey, C., Bothun, G., \& Malin, D. 
1988, \apj, 330, 634

\bibitem[Irwin \& Hatzidimitriou(1995)]{irw95}Irwin, M. \&
Hatzidimitriou, D. 1995, \mnras, 277, 1354 

\bibitem[Jerjen, Freeman \& Binggeli(2000)]{jer00}Jerjen, H., Freeman,
K. C., \& Binggeli, B. 2000, \aj, 119, 166 

\bibitem[Jerjen \& Tammann(1997)]{jer97}Jerjen, H. \& Tammann,
G. A. 1997, \aap, 321, 713

\bibitem[Kambas et al.(2000)]{kam00}Kambas, A., Davies, J. I., Smith,
R. M., Bianchi, S., \& Haynes, J. A. 2000, \aj, 120, 1316

\bibitem[Karachentsev(1996)]{kar96}Karachentsev, I. 1996, \aap, 305,
33 

\bibitem[Karachentseva \& Karachentsev(1998)]{kar98} Karachentseva,
V. E. \& Karachentsev, I. D. 1998, A\&AS, 127, 409

\bibitem[Landolt(1992)]{lan92}Landolt, A. U. 1992, \aj, 104, 340

\bibitem[Lavery \& Mighell(1992)]{lav92}Lavery, R. J. \& Mighell,
K. J.  1992 \aj, 103, 81

\bibitem[Lin et al.(1996)]{lin96} Lin, H., Kirshner, R. P., Schectman,
S. A., Landy, S. D., Oemler, A., Tucker, D. L., \& Schechter, P. L. 1996,
\apj, 464, 60

\bibitem[Loveday et al.(1992)]{lov92} Loveday, J., Peterson, B. A.,
Efstathiou, G., \& Maddox, S. J. 1992, \apj, 390, 338

\bibitem[Loveday(1997)]{lov97}Loveday, J. 1997, \apj, 489, 29

\bibitem[Marzke et al.(1994)]{mar94} Marzke, R. O., Huchra, J. P., \&
Geller, M. J. 1994, \apj, 428, 43

\bibitem[Marzke et al.(1998)]{mar98} Marzke, R. O., Da Costa, L. N.,
Pellegrini, P. S., Willmer, C. N. A., \& Geller, M. J. 1998, \apj, 503,
617

\bibitem[Mateo(1998)]{mat98} Mateo, M. 1998, \araa, 36, 435

\bibitem[Metcalfe et al.(1998)]{met98}Metcalfe, N., Ratcliffe, A.,
Shanks, T., \& Fong, R. 1998, \mnras, 294, 147

\bibitem[O'Neil, Bothun, \& Impey(1999)]{one99}O'Neil, K., Bothun,
G. D., \& Impey, C. D. 1999, \aj, 118, 1618

\bibitem[O'Neil, Bothun, \& Schombert(2000)]{one00}O'Neil, K., Bothun,
G. D., \& Schombert, J. 2000, \aj, 119, 136

\bibitem[Phillipps et al.(1998)]{phi98}Phillipps, S., Parker, Q. A.,
Schwartzenberg, J. M., \& Jones, J. B. 1998, \apj, 493, L59

\bibitem[Pritchet \& van den Bergh(1999)]{pri99}  Pritchet, C. J. \&
van den Bergh, S. 1999, \aj, 118, 883

\bibitem[Sakai et al.(1997)]{sak97}Sakai, S., Madore, B. F., Freedman,
W. L., Lauer, T. R., Ajhar, E. A., \& Baum, W. A.  1997, \apj, 478, 49

\bibitem[Sandage \& Binggeli(1984)]{san84}Sandage, A. \& Binggeli,
B. 1984, \aj, 89, 919

\bibitem[Sandage, Binggeli, \& Tammann(1985)]{san85}Sandage, A.,
Binggeli, B., \& Tammann, G. A. 1985, \aj, 90, 1759

\bibitem[Schechter(1976)]{sch76} Schechter, P. 1976, \apj, 203, 297

\bibitem[Telles \& Terlevich(1997)]{tel97}Telles, E. \& Terlevich, R. 1997,
\mnras, 286, 183

\bibitem[Thuan(1983)]{thu83}Thuan, T. X. 1983, \apj, 268, 667

\bibitem[Tonry \& Schneider(1988)]{ton88}Tonry, J. \& Schneider,
D. P., 1988, \aj, 96, 807 

\bibitem[Trentham(1998a)]{tre98a} Trentham, N. 1998a, \mnras, 294, 193

\bibitem[Trentham(1998b)]{tre98b} Trentham, N. 1998b, in Dwarf Galaxies
and Cosmology, ed. T.X. Thuan et al. (Gif-sur-Yvette: Editions
Frontieres), in press (astro-ph/9804013)  

\bibitem[Tully(1987)]{tul87}Tully, B. 1987, \apj, 321, 280

\bibitem[Tully(1988)]{tul88}Tully, R. B. 1988, Nearby Galaxies Catalogue
(Cambridge: Cambridge University Press)

\bibitem[Vader \& Chaboyer(1992)]{vad92}Vader, J. P. \& Chaboyer,
B. 1992, PASP, 104, 57

\bibitem[Vader \& Chaboyer(1994)]{vad94}Vader, J. P. \& Chaboyer,
B. 1994, \aj, 108, 1209

\bibitem[Vader \& Sandage(1991)]{vad91}Vader, J. P. \& Sandage, A.
1991, \apj, 379, L1

\bibitem[Valdes(1998)]{msc98}Valdes, F. 1998, Guide to the NOAO Mosaic
Data Handling System, IRAF MSCRED V2.0

\bibitem[Valotto, Moore, \& Lambas(2000)]{val00}Valotto, C. A., Moore, B., \&
Lambas, D. G. 2000, \apj, in press (astro-ph/0009230)

\bibitem[Zabludoff \& Mulchaey(2000)]{zab00}Zabludoff, A. I. \&
Mulchaey, J. S. 2000, \apj, 539, 136

\bibitem[Zaritsky et al.(1997)]{zar97}Zaritsky, D., Nelson, A.,
Dalcanton, J., \& Gonzalez, A. 1997, \apj, 480, L91

\bibitem[Ziegler \& Bender(1998)]{zie98}Ziegler, B. L. \& Bender,
R. 1998, \aap, 330, 819

\bibitem[Zucca et al.(1997)]{zuc97} Zucca, E., et al.
1997, \aap, 326, 477

\bibitem[Zwicky et al.(1968)]{zwicky}Zwicky, F., et al. 1968,
Catalogue of Galaxies and Clusters of Galaxies (Pasadena: California
Institute of Technology) 

\end{thebibliography}
\end{document}